\documentclass[aip,jcp,amsmath,amssymb,preprint]{revtex4-1}

\usepackage{graphicx}
\usepackage{dcolumn}
\usepackage{bm}
\usepackage{comment}
\usepackage{xcolor}
\usepackage{soul}

\begin{document}


\title{A corresponding-state framework for the structural transition of supercritical fluids across the Widom delta}

\author{Tae Jun Yoon}
\affiliation{School of Chemical and Biological Engineering, Institute of Chemical Processes, Seoul National University, Seoul 08826, Republic of Korea}
 
\author{Min Young Ha}
\affiliation{School of Chemical and Biological Engineering, Institute of Chemical Processes, Seoul National University, Seoul 08826, Republic of Korea}

\author{Won Bo Lee}
\email{wblee@snu.ac.kr}
\affiliation{School of Chemical and Biological Engineering, Institute of Chemical Processes, Seoul National University, Seoul 08826, Republic of Korea} 

\author{Youn-Woo Lee}
\email{ywlee@snu.ac.kr}
\affiliation{School of Chemical and Biological Engineering, Institute of Chemical Processes, Seoul National University, Seoul 08826, Republic of Korea}
 
\date{\today}

\begin{abstract}
This work proposes a classification algorithm based on the radical Voronoi tessellation to define the Widom delta, supercritical gas-liquid coexistence region, of polyatomic molecules. In specific, we use a weighted mean-field classification method to classify a molecule into either gas-like or liquid-like. Classical percolation theory methods are adopted to understand the generality of the structural transition and to locate the Widom delta. A structural analysis on various supercritical fluids shows that the proposed method detects the influence of the attractive interaction on the structural transition of supercritical fluids. Moreover, we demonstrate that the supercritical gas-liquid coexistence region of water overlap with the ridges of the response function maxima. From the pressure-temperature relation, a three-parameter corresponding state theorem is derived, which states that the fraction of gas-like molecules of a substance is equal to that of another if their reduced pressure, reduced temperature and the critical compressibility factor are the same.
\end{abstract}

\maketitle

\section{\label{sec:level1}Introduction}
Supercritical fluid, a state of matter beyond the liquid-gas critical point, is regarded as one of the most promising alternatives to conventional liquid solvents used in chemical processes~\cite{nalawade2006supercritical,perrut2000supercritical}. One of the useful physical characteristics for the industrial application is its great tunability: small changes in the thermodynamic conditions result in the continuous yet substantial adjustment of thermophysical properties and thermodynamic response functions~\cite{vega2017perspectives,yoon2017molecular}. It is currently believed that this tunability is deeply related to the inhomogeneous nature of supercritical fluids, refuting the conventional point of view which sees it as a monotonic and homogeneous state of matter.

Thus, a considerable amount of theoretical and experimental works were devoted to understanding the anomalous behavior of supercritical fluids. In scattering and spectroscopic measurements, a large density fluctuation was observed, which imply the existence of the structural inhomogeneities~\cite{nishikawa1996small,pipich2018densification,tucker1999solvent}. Anomalous behaviors of the response functions and transport properties in the vicinity of the critical point also supported the existence of two different states in the supercritical region~\cite{simeoni2010widom,brazhkin2011widom,banuti2015crossing}, where the continuous crossover between these states is reminiscent of subcritical vapor-liquid transition. In order to capture this ``supercritical gas-liquid transition'', or the continuous structural transition and the resultant thermodynamic anomalies beyond the critical point, the following thermodynamic methods were suggested.

First, thermodynamic response functions and the correlation length were used to locate the thermodynamic states where the supercritical gas-liquid transition occurs. Simeoni et al. proposed that the Widom line, which was originally defined as a loci of the maximum correlation length, distinguishes gas-like and liquid-like behaviors of supercritical fluids.~\cite{simeoni2010widom}. Banuti et al. suggested that the line of heat capacity maxima, one of the so called pseudo-boiling lines, follows the three-parameter corresponding state principle~\cite{banuti2017similarity}. This approach, which assumes the existence of a single line that demarcates the supercritical fluid region, has also been discussed to understand the anomalous behavior of pure supercritical fluids~\cite{imre2012pseudocritical,imre2015anomalous,gallo2014widom} and their mixtures (supercritical liquid-gas boundary)~\cite{raju2017widom}, and extended to understand the liquid-liquid transition~\cite{dey2013information,fuentevilla2006scaled,abascal2010widom}. However, the response-function based separation of the supercritical region has shown its limits. First, the magnitude of the correlation length and response function maxima rapidly decays and disappears when far from the critical temperature~\cite{brazhkin2011widom}. Second, Widom line and the pseudo-boiling lines only overlap near the critical point; they diverge from each other departing from the critical point\cite{brazhkin2011widom,may2012riemannian,brazhkin2014true,fomin2015thermodynamic}. Third, the location of the response function maxima shows a path-dependence~\cite{fomin2011inversion,fomin2015thermodynamic,schienbein2018investigation}. The lines of the response function maxima on isobars and isotherms do not overlap with each other. From these remarks, Schienbein and Marx~\cite{schienbein2018investigation} recently addressed that no unique separating ``line'' can be deduced from the response functions.

Independent from these response-function based methods, the structural characteristics of the supercritical fluid have also been studied, which can be dated back to the works of Bernal~\cite{bernal1959geometrical,bernal1964bakerian}. In the spirit of the Bernal's work, Finney~\cite{finney2014renaissance}, Woodcock, and Heyes~\cite{heyes2013critical} located the percolation transition loci based on the rigidity calculation of the monatomic fluid. The pair correlation function integrals were also used to understand the nature of the inhomogeneous structure of near-critical fluids~\cite{tucker1999solvent,song2000intermolecular,skarmoutsos2009effect,ploetz2017fluctuation}. Several molecular-level classification schemes have also been proposed to define the gas-like and the liquid-like structures in supercritical fluids. Traditionally, the fixed-distance cutoff methods~\cite{stillinger1963rigorous,idrissi2013characterization,heyes1988percolation,vskvor2009percolation} were used to analyze the inhomogeneous structure of supercritical fluids based on the cluster analysis, but the fixed cutoff methods were susceptible to the selection of the cutoff radius~\cite{vskvor2009percolation}. Hence, a series of recent works have used the Voronoi tessellation as a tool to define the local structure of supercritical fluids~\cite{idrissi2010local,idrissi2011heterogeneity,vskvor2011percolation}. For instance, Ovcharov et al. proposed a method for phase identification (MPI) based on the Voronoi tessellation to define liquid, gas, and surface atoms in the two-dimensional system~\cite{ovcharov2017particle}. 

Our recent works have focused on the application of the Voronoi tessellation to analyze the structural transition of monatomic supercritical fluids. Implementing machine learning~\cite{ha2018widom} and probabilistic classification~\cite{yoon2018probabilistic} strategies, we have successfully identified distinct `liquid-like' and `gas-like' structures coexisting in supercritical fluid, and have located regions where the two microstates coexist, which we termed the `Widom delta'. This approach is based on the theory of fluid polyamorphism~\cite{anisimov2018thermodynamics} or the heterogeneous-homogeneous fluctuation~\cite{nilsson2015structural}, where the macroscopic anomalies of fluid phase is understood from the interplay of the two microscopic structures.

This geometric definition of the supercritical gas-liquid coexistence region, called the Widom delta, has shown the following advantages compared to the conventional approaches. First, the boundaries of the Widom delta agree with those obtained from the rigidity calculation~\cite{heyes2013critical}. They are also similar to the coexistence lines detected from the MPI algorithm~\cite{ovcharov2017particle}. Second, neither path-dependence nor termination of the coexistence region was observed. Thus, a contradiction which comes from the path-dependence of the response functions does not occur, and the coexistence region does not vanish even in the deeply supercritical conditions far from the critical point~\cite{yoon2018probabilistic}. Third, the algorithm depends only on the critical density as a classification criterion, which is an intrinsic property of a fluid. Since selecting a proper cutoff distance comes with a certain degree of arbitrariness, a classification scheme without the cutoff distance is meritorious. Lastly, defining a coexistence ``region'' instead of the ``ridges of the response function maxima'' for the demarcation of the supercritical fluid phase is adequate to describe the continuous structural variation of supercritical fluids~\cite{hestand2019mid}; no discontinuous (first-order) phase transition occurs above the critical point.
 
This work aims to extend the classification method, originally demonstrated in the simple monatomic Lennard-Jones fluid  system~\cite{yoon2018probabilistic}, to examine the structural transition of polyatomic supercritical fluids and to discover their general structural characteristics. We first compute the fraction of gas-like molecules ($\Pi_{gas}$) by classifying molecules into gas-like and liquid-like applying the geometric procedure based on the radical Voronoi (Laguerre) tessellation. The characteristics of the gas-like and liquid-like structures are examined based on the percolation theory. The classical percolation theory enables us to shed light on the generality of the structural evolution based on the fraction of gas-like molecules. Specifically, we locate the supercritical gas-liquid coexistence region including the supercritical gas-liquid boundary\footnote{The term supercritical gas-liquid boundary, which we introduced in previous works, or any other terminologies used by previous works (supercritical liquid-gas boundary \cite{raju2017widom}, pseudo-boiling line \cite{banuti2017similarity} and hypercritical line\cite{bernal1964bakerian}) do not exactly match with the subcritical gas-liquid transition which involves both orientational (topological) and spatial variation of the system. Rather, these terms in addition to `gas-like' or 'liquid-like' regard the large variation of spatial order observed across the critical density as the remark of the pseudo-boiling.}, which is defined as the thermodynamic states where $\Pi_{gas}$ becomes 0.5 following the philosophy of the fluid polyamorphism \cite{anisimov2018thermodynamics}, and the percolation transition lines from the finite-size scaling analysis.

We further demonstrate the generality of the supercritical gas-liquid coexistence region on the phase diagrams. The density-temperature diagram shows that the supercritical gas-liquid coexistence region is quasi-universal; the coexistence regions of non-polar and polar substances almost overlap with each other except strongly hydrogen-bonded substances. A comparison between the supercritical gas-liquid coexistence region of water and its lines of the response function maxima reveals that the supercritical gas-liquid coexistence region defined in this work agrees well with most of these demarcation lines. On the pressure-temperature diagrams, we show that the deltoid shape of the supercritical gas-liquid coexistence region is preserved for all substances. From simple relationships based on the compressibility factors, we further demonstrate that supercritical gas-liquid transition of various substances follows the extended corresponding state principle.

\section{Computational details}
\subsection{Molecular Dynamics (MD) simulations}
Ten substances are selected to investigate the general characteristics of the Widom delta. The interaction parameters for methane, ethane, carbon dioxide, nitrogen, oxygen, and ethylene oxide are modeled using the Transferable Potentials for Phase Equilibria (TraPPE) forcefields~\cite{martin1998transferable,potoff2001vapor,zhang2006direct,ketko2008development}. For water and methanol, the potential parameters are adopted from the TIP4P/2005~\cite{abascal2005general} and OPLS/2016~\cite{gonzalez2016new} forcefields, respectively. The interaction parameters for ammonia are obtained from work by Eckl et al.~\cite{eckl2008optimised}. The critical points of these substances are estimated based on the flat top proposal. According to the flat top proposal~\cite{heyes2013critical}, the rigidity and its first derivative of a system become zero at the critical point ($d{\rho}/dp=0$ and $d{\rho^2}/dp^2=0$).
\begin{table}
	\caption{Critical constants estimated based on the flat top proposal.} 
	\begin{ruledtabular}
		\centering
		\begin{tabular}{lcccc}
			Substances & $T_{c}$ [K] & $p_{c}$ [bar] & $\rho_{c}$ [$kg/m^{3}$] & $z_{c}$\\
			\hline
			$\mbox{Ar}$ & 164.31 & 63.93 & 471.30 & 0.397\\
			$\mbox{N}_{2}$ & 129.97 & 38.61 & 291.91 & 0.343\\
			$\mbox{O}_{2}$ & 158.02 & 58.39 & 398.78 & 0.357\\
			$\mbox{CH}_{4}$ & 199.52 & 56.55 & 147.21 & 0.371\\
			$\mbox{C}_{2}\mbox{H}_{6}$ & 312.08 & 60.09 & 195.52 & 0.356\\
			$\mbox{CH}_{2}\mbox{OCH}_{2}$ & 470.68 & 97.16 & 358.11 & 0.305\\
			$\mbox{CO}_{2}$ & 311.99 & 86.50 & 452.35 & 0.324\\
			$\mbox{NH}_{3}$ & 414.60 & 132.42 & 215.09 & 0.304\\
			$\mbox{H}_{2}\mbox{O}$ & 654.10 & 178.90 & 284.23 & 0.208\\		
			$\mbox{CH}_{3}\mbox{OH}$ & 537.07 & 89.14 & 247.32 & 0.259\\
		\end{tabular}   
	\end{ruledtabular}
	\label{table:critical-point}
\end{table}

To calculate the rigidity and its first derivative, we perform NVT simulations~\cite{plimpton1995fast} in the vicinity of the critical points reported in the earlier works. The number of molecules is 1,000. The timestep is 1 fs. The interatomic potentials are truncated at the cutoff radius of $16.0$ $\mbox{\AA}$ and the analytic tail correction term is added. For polyatomic species, the Lorentz-Berthelot mixing rules are used to model the interaction between different atoms:
\begin{equation}
\sigma_{ij}=\frac{\sigma_{ii}+\sigma_{jj}}{2}; \epsilon_{ij}=\sqrt{\epsilon_{ii}\epsilon_{jj}}.
\end{equation}
The systems are equilibrated for 100,000 steps, and the system pressures are collected every step during 1,000,000 steps. By fitting the cubic equations, the inflection points at different isotherms were obtained. The critical temperature is then calculated as the temperature where $d\rho/dp$ at the inflection point becomes zero. Then, the critical density and pressure are computed. Table \ref{table:critical-point} shows the critical points of the substances calculated in this manner. After the critical point of each substance is estimated, we perform NVT simulations at $T_{r}=T/T_c=1.00-3.70$ and $\rho_{r}=\rho/\rho_c=0.2-2.4$ with the same cutoff radius and the number of molecules. The systems are equilibrated for 100,000 steps, and the system configurations are collected every 10,000 steps during the production run (1,000,000 steps). The pressure data are sampled every step. We also simulate supercritical argon at $T_{r}=3.70$ and $\rho_{r}=0.75-1.17$ to examine the system size effect and to carry out the finite-size scaling analysis. The numbers of molecules are $N=1,000, 2,000, 4,000, 8,000,$ and $16,000$. The simulations for the finite-size scaling analysis include 100,000 steps of equilibration run followed by 500,000 steps of data production. The trajectories are collected every 1,000 steps. Besides, we also perform the subcritical NVT simulations of the Lennard-Jones particles ($\sigma=1.0$ and $\epsilon=1.0$) to revalidate the classification strategy. These results can be found in the supplementary material.

\subsection{Classification strategy}
\begin{figure*}
	\includegraphics[width=0.6\textwidth]{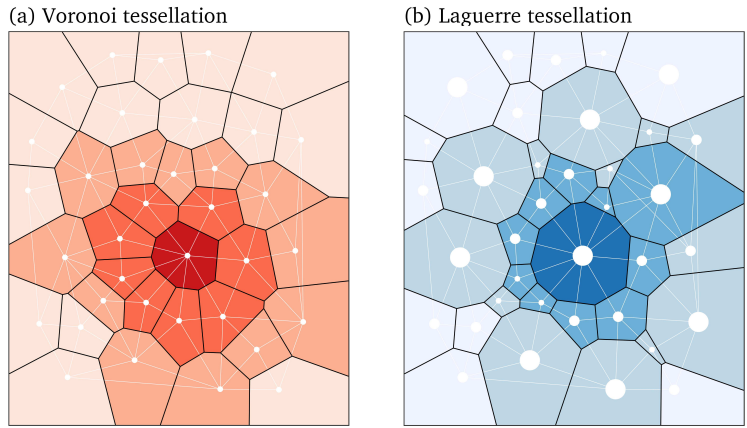}
    \caption{Two-dimensional representation of (a) simple Voronoi tessellation and (b) radical Voronoi (Laguerre) tessellation of the particles at the same positions. In the Laguerre tessellation, the power distance is used to partition the space into per-atom cells: Note that spheres with larger radii claim larger area in the space. The color scheme denotes the weighted mean-field strategy to classify a monatomic molecule based on the Voronoi network. To classify a central molecule (dark cell), local densities of molecules up to its second neighbors are averaged with a weight based on the chemical distance.
    \label{fig:explanation}}
\end{figure*}

A probabilistic classification algorithm on the LJ monatomic fluid~\cite{yoon2018probabilistic} is extended to the polyatomic substances as follows. Since the atomic radii of polyatomic molecules are different, we adopt the radical Voronoi tessellation (Laguerre tessellation) to partition the space into per-atom cells and define the local density of a molecule. In the radical Voronoi tessellation, a configuration of $N$ polydisperse spheres is partitioned into $N$ cells based on the power distance between each other~\cite{rycroft2009voro++}. The power distance of a point with respect to a sphere is defined as the squared length of the tangential line from the point to the sphere, hence is dependent on the radius of the sphere (Fig.~\ref{fig:explanation}b). The Laguerre cell of a sphere S is then defined as the set of points that are 'closer' to the S than any other spheres, in terms of the power distance metric. Note that the radical Voronoi tessellation is identical to conventional Voronoi tessellation when the spheres in the system are monodisperse (Fig.~\ref{fig:explanation}a).

Thus, a system with $N$ molecules which has $n_a$ atoms each has $n_{a}N$ Laguerre cells obtained from radical Voronoi tessellation. We adopt the atomic radii as the LJ distance parameter, $\sigma_{jj}$ for atom type $j$ except water and methanol. Since the atomic radii of hydrogen atoms are given as zero in OPLS/2016 and TIP4P/2005 models, their configurations are partitioned into simple Voronoi cells using the oxygen atoms as the seeds. The local volume of $i^{\rm{th}}$ molecule is defined as the sum of Laguerre cells of its constituting atoms, and its local density is defined as the reciprocal of its local volume:
\begin{equation}
	\rho_{i}=\frac{1}{\sum_{j=1}^{n_{a}}v_{i}^{j}}
\end{equation}
The neighbors of the $i^{\rm{th}}$ molecule is defined as the molecules of which atomic Laguerre cells share faces with those of the $i^{\rm{th}}$ molecule, and the second neighbors are defined as the neighbors of neighbors.

After the local density data of the molecules are obtained, the following weighted mean-field strategy is used to determine whether a molecule is gas-like or liquid-like. This weighted mean-field strategy exploits the chemical distance to give a weight on the central particle. When the weighted mean-field local density of the $i^{\rm{th}}$ molecule is supposed to be obtained, it is given as:
\begin{equation}
	\bar{\rho_{i}}=\frac{1}{N_{i}+1}\sum_{j\in\rm{NN}(\it{i})}\left(\frac{1}{N_{j}+1}\sum_{k\in\rm{NN}(\it{j})}\rho_{k}\right)
	\label{eqn:weighted-mean-field}
\end{equation}
where $N_{m}$ is the number of neighbors of the $m^{th}$ molecule, and $\rm{NN}(\it{m})$ is the set of neighbors of the $m^{\rm{th}}$ molecule. For instance, when counting cells in the two-dimensional systems shown in Fig. \ref{fig:explanation}, the local density of the central atom (dark blue cell) which has ten nearest neighbors is counted eleven times. The local densities of the nearest neighbors are counted four times. The second nearest neighbors are counted either once or twice. If the second nearest neighbor shares two faces with two first nearest neighbors, it is counted twice; otherwise, it is counted once. By averaging the local densities with these weights, one can obtain the weighted mean-field local density of the central atom. The same procedure is applied for the three-diemensional systems. Finally, if this averaged local density is higher than the threshold density, the $i^{\rm{th}}$ molecule is classified as liquid-like. Otherwise, it is gas-like. 

After the weighted mean-field classification is carried out, the gas-like fraction of a system is calculated as:
\begin{equation}
    \Pi_{\rm{gas}}=\frac{N_{\rm{gas}}}{N_{\rm{gas}}+N_{\rm{liq}}}; \Pi_{\rm{liq}}=1-\Pi_{\rm{gas}}
    \label{eqn:pigas-definition}
\end{equation}
where $N_{\rm{gas}}$ and $N_{\rm{liq}}$ mean the numbers of gas-like and liquid-like molecules, respectively. This classification result depends on the choice of the threshold density. When various threshold densities are applied to subcritical vapor and liquid phases, we can find that the following result is obtained when the critical density is selected as the threshold density.
\begin{equation}
    \left\lbrace
    \begin{aligned}
    \Pi_{\rm{gas}}&=1.0\quad(\rm{vapor}) \\
    \Pi_{\rm{gas}}&=0.0\quad(\rm{liquid})
    \end{aligned}
    \right.
\end{equation}
When a threshold density lower than the critical density is used to classify molecules, a considerable amount of vapor molecules are classified as liquid-like ones, and vice versa (see the Supplementary material for the results). Taking into account that the saturated vapor/liquid systems should be clearly distinct from each other, we chose the critical density as the classification density for all substances in this work.

\subsection{Percolation analysis}
In our last work, we estimated the percolation transition densities $\rho_{pb}$ and $\rho_{pa}$ of the monatomic LJ fluid from the gas-like fraction curve. They were defined as the densities where $\Pi_{\rm{gas}}$ starts to change steeply.
\begin{equation}
    \rho_{pa}=\frac{\log(a)+2}{b}; \rho_{pb}=\frac{\log(a)-2}{b}
    \label{eqn:old-def-percolation}
\end{equation} 
where $a$ and $b$ are the parameters obtained by fitting a sigmoid equation to the gas-like fraction data. Here, $\rho_{pa}$ was called the percolation transition density of the available volume and $\rho_{pb}$ was the percolation transition density of the bonded clusters (liquid droplets)~\cite{heyes2013critical}. This work estimates $\rho_{pa}$ and $\rho_{pb}$ in a more elaborate manner based on the classical percolation theory. Percolation theory states that a spanning (infinite) cluster connected across the periodic boundary emerge as the occupation probability of a particle increases~\cite{stauffer2014introduction}. Here, the term ``cluster'' is not limited to denote the high-density (liquid-like) particles. Instead, it is mathematically defined as a group of neighboring atoms whose identity are same to each other following the classical percolation theory~\cite{stauffer2014introduction}. We refer to the cluster of gas-like and liquid-like molecules as `bubble` and `droplet', respectively.

The occupation probability at which an infinite cluster appears is called the percolation threshold. To examine the percolation behavior, we apply the clustering algorithm proposed by Stoll~\cite{stoll1998fast} to the Voronoi network. In this algorithm, two molecules are regarded to be connected if their classification results are identical and they are the nearest neighbors to each other. Hence, two molecules distant from each other are included in the same cluster if there exists a string of molecules of the same type linking them. The algorithm first counts the independent clusters without regard for the periodic boundary conditions. Second, it tests whether a cluster is infinite or not. The algorithm deems a cluster as an infinite one if one of its molecules at the periphery of the simulation box is connected to another one at the opposite side. After the infinity test, the independent clusters obtained in the first step are connected if they are linked across the periodic boundary conditions.

After the infinity test, we obtain the percolation probability ($p_{inf}$), the probability to find an infinite cluster (gas-like bubble or liquid-like droplet) by dividing the number of configurations which contain at least one infinite cluster by the number of total configurations at the given condition. Since each configuration is independent from each other, the confidence interval (CI) of the percolation probability can be defined from the binomial distribution.
\begin{equation}
    CI = \frac{2}{N_{conf}}\sqrt{p_{inf}N_{conf}\left(1-p_{inf}\right)}
    \label{eqn:binomial-confidence}
\end{equation}
where $N_{conf}$ is the number of configurations (snapshots). A finite-size scaling analysis is then carried out to determine $\Pi_{thr}^{\infty}$ in the following manner~\cite{stauffer2014introduction}. We first fit the following sigmoidal equation to $p_{inf}$ data.
\begin{equation}
    p_{inf}^{i}=\left[1+\exp(a\Pi_{i}+b)\right]^c
    \label{eqn:sigmoid-pinf}
\end{equation}
where $i$ is the type of the molecule ($i=\mbox{gas}, \mbox{liq}$). Then, the average percolation threshold ($\Pi_{i}^{av}$) and the transition region width ($\Delta$) are computed as:
\begin{align}
&\Pi_{i}^{av}=\int\Pi_{i}\left(\frac{dp_{inf}^{i}}{d\Pi_{i}}\right)d\Pi_{i}\\
&\Delta_{i}^2=\int(\Pi_{i}-\Pi_{i}^{av})^2\left(\frac{dp_{inf}^{i}}{d\Pi_{i}}\right)d\Pi_{i}
\label{eqn:piav-delta-def}
\end{align}
The percolation threshold for an infinite system ($\Pi_{i}^{\infty}$) is then obtained as an intercept of the following linear equation.
\begin{equation}
    \Pi_{i}^{av}=a\Delta_{i}+\Pi_{i}^{\infty} 
    \label{eqn:finite-size-scaling}
\end{equation}
where $a$ and $\Pi_{i}^{\infty}$ are fitting parameters.

\section{Results and Discussion}
\subsection{Molecular-level classification}
\begin{figure}
	\includegraphics{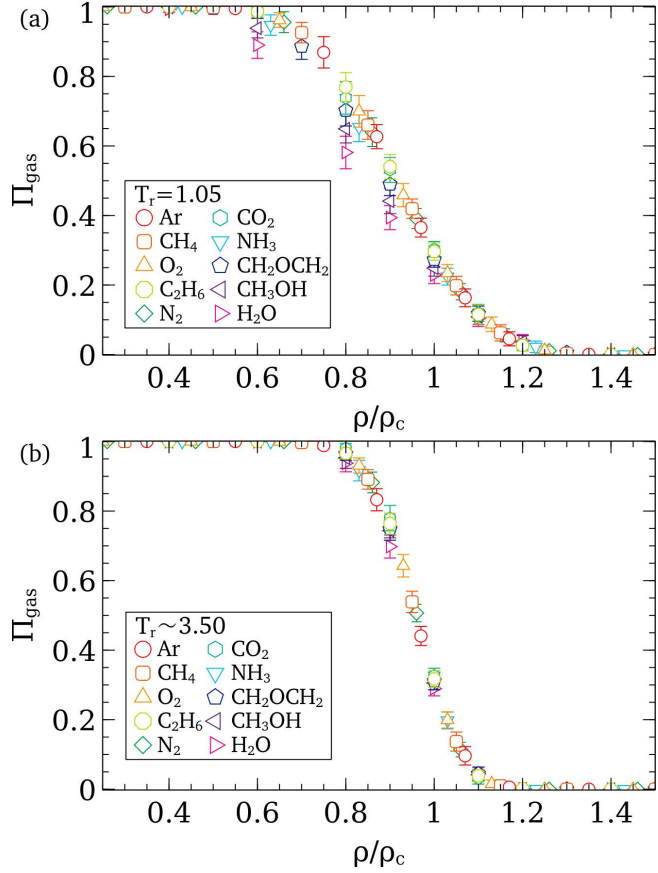}
    \caption{Dependence of the gas-like fraction ($\Pi_{\rm{gas}}$) on the bulk density at (a) $T_{r}=1.05$ and $T_{r}\sim3.50$. The fraction of gas-like molecules decreases from one to zero as the density increases. At $T_{r}=1.05$, the gas-like fractions of strongly polar substances start to decrease at lower densities than non-polar substances. As the temperature increases, the gas-like fraction curves of all substances collapse to a single line. Error bars indicate the standard deviation of $\Pi_{\rm{gas}}$ of the systems.
    \label{fig:pigascurve}}
\end{figure}
Fig. \ref{fig:pigascurve} shows the fraction of gas-like molecules at $T_{r}=1.05$ and $T_{r}=3.50$. As observed in our earlier work on the monatomic LJ fluid, the dependence of gas-like fraction on the bulk density could be expressed by an inverse sigmoid function based on the fluid polyamorphism formulation~\cite{anisimov2018thermodynamics,yoon2018probabilistic,ha2018widom}. In the theory of fluid polyamorphism, the structural transition between two different amorphous structures is regarded as a reaction.
\begin{equation}
A(\mbox{gas}) \rightleftarrows A(\mbox{liquid}) 
\end{equation}
The equilibrium constant $K_{eq}$ is then calculated as $K_{eq}=\Pi_{liquid}/\Pi_{\rm{gas}}=\exp(-\beta{\Delta}G^{\ddagger})$ where $\beta$ is the thermodynamic beta ($1/k_{B}T$) and ${\Delta}G^{\ddagger}$ is the Gibbs energy difference between the amorphous structures. By assuming that ${\Delta}G^{\ddagger}$ is proportional to the bulk density, the following equation can be derived.

\begin{equation}
\Pi_{\rm{gas}}=1-\frac{1}{1+a\exp(-b\rho_{r})}
\label{eqn:pigas-sigmoid}
\end{equation}

\begin{figure*}
	\includegraphics[width=\textwidth]{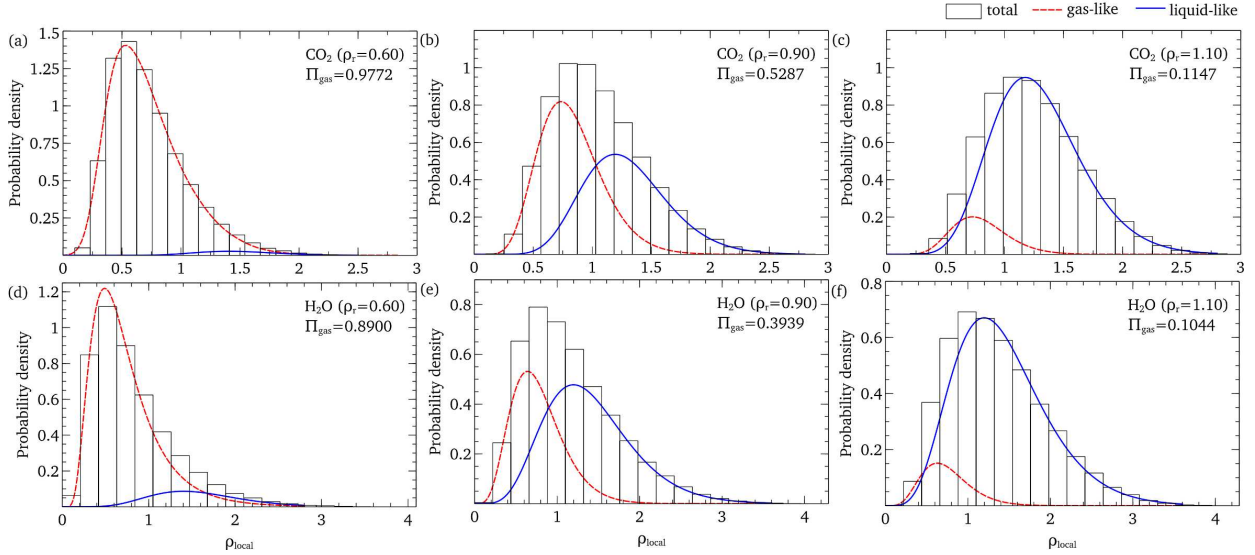}
    \caption{Local density distributions of carbon dioxide (a-c) and water (d-f) at $T_{r}=1.05$. The fraction of liquid-like molecules of water is always greater than that of carbon dioxide in the supercritical gas-liquid coexistence region, which substantiates the strong local density augmentation induced by the attractive interactions.
    \label{fig:locdendist}}
\end{figure*}

At $T_{r}=1.05$, the inverse sigmoid functions of polar substances do not agree with those of non-polar/weakly polar substances (Fig. \ref{fig:pigascurve}a). $\Pi_{\rm{gas}}$ of polar substances start to decrease at lower densities than those of non-polar ones. When the temperature increases, the inconsistency of $\Pi_{\rm{gas}}$ curves of different substances diminishes (Fig. \ref{fig:pigascurve}b). This temperature dependence substantiates that the supercritical gas-liquid transition across the Widom delta is largely influenced by the competition between the attractive force and the kinetic energy. Near the critical point, the magnitude of the kinetic energy is comparable to the attraction due to its neighbors. Hence, the magnitude of the intermolecular attraction matters. As the temperature increases, a significant proportion of molecules possess kinetic energy that can overcome the attractive forces among the molecules. As a result, the supercritical gas-liquid transition density increases, and the discrepancy of the gas-like fraction curves among the substances decreases. This competitive relationship between the attractive energy and the kinetic energy can also be understood by examining the compressibility factor ($z{\equiv}p/\rho{k_{B}}T$). When the attractive interaction dominates, $z$ becomes smaller than unity. As shown in Table \ref{table:critical-point} and the thermodynamic properties given in the supplementary information, the compressibility factors of polar substances at the critical point are lower than the non-polar molecules. As the density increases, they decrease. On the contrary, at higher temperatures far from the critical point ($T_{r}>2$), $z$ becomes larger than one for all materials and increases as the density increases. These results indicate that the influence of the attractive interaction on the structural transition diminishes. 

It should also be noted that the $\Pi_{\rm{gas}}$ curves of different species at $T{\sim}T_{c}$ only overlap with each other at high densities. This result indicates that the Weeks-Chandler-Andersen (WCA) liquid perturbation theory~\cite{weeks1971role} cannot be used to analyze the structural characteristics of low-density and near-critical fluid~\cite{ben2004reformulation}. As Toxv{\ae}rd stated~\cite{toxvaerd2015role}, the equilibrium structure of the gas and the near-critical fluid is strongly affected by the presence of the attractive interaction.

\subsection{Structural analysis}
We further characterize the structural characteristics based on the local density distributions and the percolation theory. Fig. \ref{fig:locdendist} shows the local density distributions of carbon dioxide and water at $T_{r}=1.05$. The local density distributions of liquid-like molecules are more symmetric than those of gas-like ones. Compared to the local density distributions of carbon dioxide, those of water at the same reduced conditions (Fig. \ref{fig:locdendist}d-f) show stronger local density augmentation. Moreover, the distributions of water are broader, which means that more high-density molecules exist as a result of the strong intermolecular interactions ($z_{\rm{CO_2}}>z_{\rm{H_{2}O}}$). This result indicates that the classification method proposed in this work also detects the effect of the strong polar forces on the local density augmentation as the integration of the pair correlation function did~\cite{yoon2017monte,song2000intermolecular,skarmoutsos2009effect}.
\begin{figure*}
    \centering
    \includegraphics[width=\textwidth]{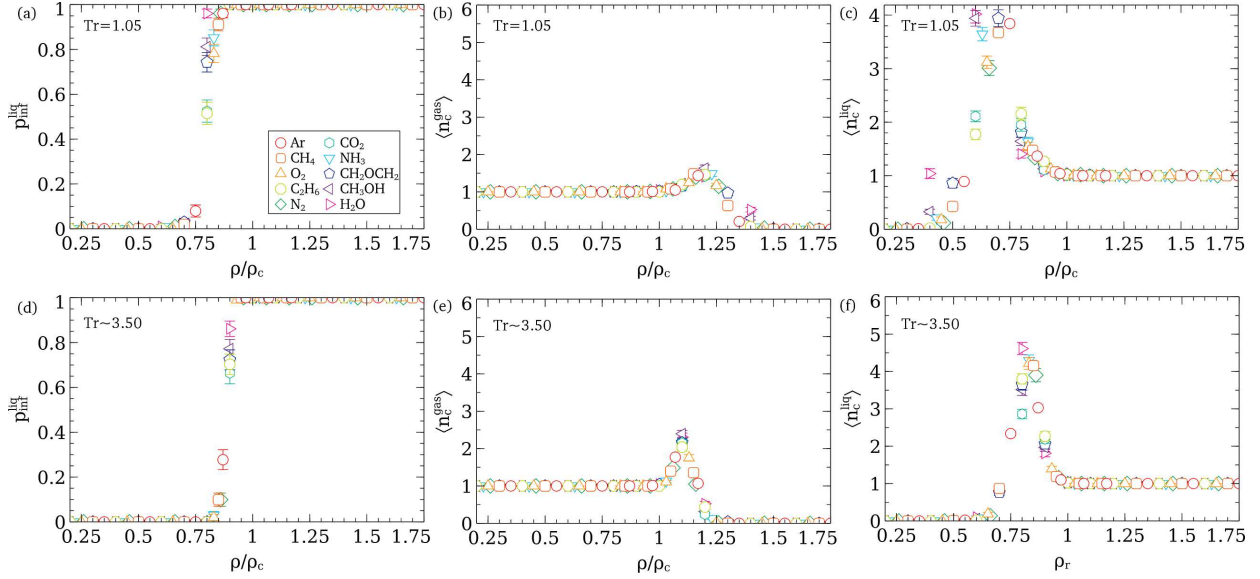}
    \caption{(a) The probability of finding an infinite liquid-like droplet in a configuration and the number of independent (b) gas-like bubbles and (c) liquid-like droplets at $T_{r}=1.05$. (d) The probability of finding an infinite liquid-like droplet in a configuration and the number of independent (e) gas-like bubbles and (f) liquid-like droplets at $T_{r}\sim3.50$. As the temperature increases, the absolute slope of $p_{inf}$ curves increases. The maximum number of independent droplets and/or bubbles also increases due to thermal agitation. The confidence intervals of $p_{inf}^{state}$ are calculated based on an assumption that all configurations are independent from each other. The standard errors of the mean number of droplets (liquids) are obtained by assuming that the sample standard deviation is equal to that of the entire population.}
    \label{fig:structural-characteristics}
\end{figure*}
Fig. \ref{fig:structural-characteristics}a-c shows the dependence of $p_{inf}$ and ${\langle}n_{c}\rangle$ on the bulk density at $T_{r}=1.05$. Here, ${\langle}n_{c}\rangle$ is the number of independent gas-like bubbles or liquid-like droplets. At $T_{r}=1.05$, $p_{inf}^{liq}$ of the substances except strongly polar molecules collapse to a single line. $p_{inf}^{liq}$ of water and methanol starts to increase at lower densities than the other fluids. The $p_{inf}^{gas}$ curves show an opposite dependence on the bulk density (see the supplementary material). At high temperatures ($T_{r}\sim3.50$), $p_{inf}^{liq}$ and $p_{inf}^{gas}$ of all substances collapse to single lines (Fig. \ref{fig:structural-characteristics}d). This temperature dependence agrees with the behavior of $\Pi_{\rm{gas}}$ shown in Fig. \ref{fig:pigascurve}. The steepness of $p_{inf}$ curves became high like the WCA fluids~\cite{yoon2018probabilistic}.

The narrower and higher peaks of ${\langle}n_{c}\rangle$ at high temperatures (Fig.~\ref{fig:structural-characteristics}e and f) reflect that the thermal agitation prevents the formation of large bubbles and/or droplets far from the critical point. In near-critical fluids, the inhomogeneous fluid structure fluctuates slowly~\cite{goodyear1999glass}. As the temperature increases, the thermal fluctuation becomes so vigorous that the inhomogeneous fluid structure cannot remain stable. As a result, the mean bubble (droplet) size decreases and the maximum number of bubbles (droplets) increases. It can also be observed that the maximum of $\langle{n_{c}^{gas}}\rangle$ is always lower than that of $\langle{n_{c}^{liq}}\rangle$. This comes from the fact that the mean local density of liquid-like molecules is higher than that of gas-like ones. Due to the large volume, an infinite gas-like bubbles can be easily formed compared to liquid-like droplets.

The similar dependence of $p_{inf}$ and $\Pi_{\rm{gas}}$ on the temperature indicates that $\Pi_{\rm{gas}}$ can work as an order parameter to describe the supercritical gas-liquid coexistence region; $p_{inf}^{liq}$ and $p_{inf}^{gas}$ curves of all substances at all temperatures collapse to single lines when $\Pi_{\rm{gas}}$ is used as an order parameter (see the supplementary material). Therefore, we conduct the finite-size scaling analysis to calculate the percolation threshold ($\Pi_{i}^{\infty}$).  
\begin{figure}
    \centering
    \includegraphics{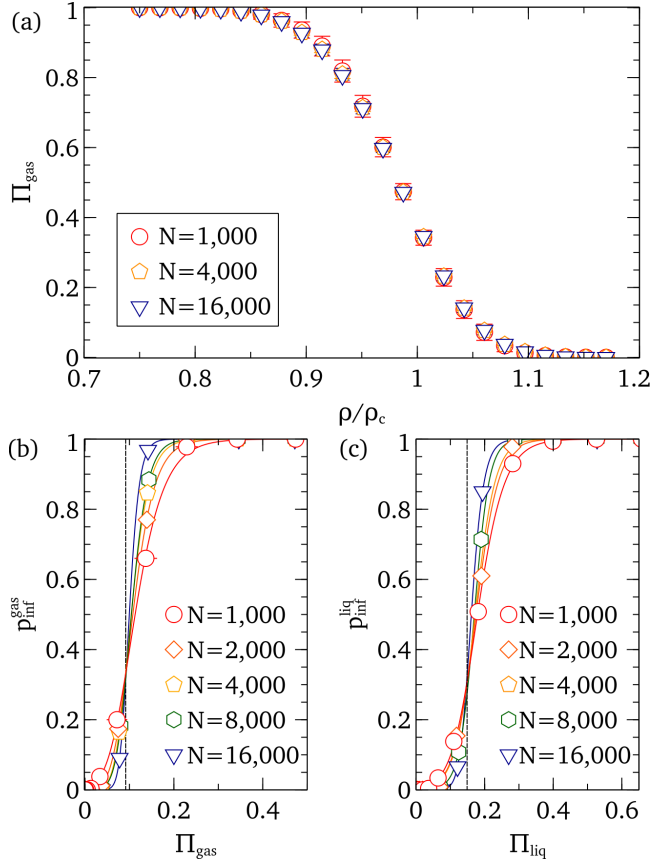}
    \caption{(a) Effect of the system size on the classification result and the finite-size scaling results of (b) gas-like bubbles and (c) liquid-like droplets of supercritical argon. No significant system size effect was observed for the classification results. The dotted lines denote the percolation thresholds for infinite systems obtained from the finite-size scaling analysis. Error bars in (b) and (c) are behind the symbols.}
    \label{fig:finite-size-scaling}
\end{figure}
Fig. \ref{fig:finite-size-scaling}a demonstrates that no significant finite-size effect exists for $\Pi_{\rm{gas}}$. This result again supports our idea that the fraction of gas-like molecules can be used as a robust parameter to define the supercritical gas-liquid coexistence region. Fig. \ref{fig:finite-size-scaling}b shows the dependence of $p_{inf}^{gas}$ on $\Pi_{\rm{gas}}$. Unlike $\Pi_{\rm{gas}}$, $p_{inf}^{i}$ largely depends on the system size. As the number of molecules in a system increases, the $p_{inf}^{gas}$ curve becomes like a step function. $p_{inf}^{liq}$ shows a similar behavior as a function of $\Pi_{\rm{liq}}$ (Fig. \ref{fig:finite-size-scaling}c). By applying Eqn.~(\ref{eqn:piav-delta-def}) and Eqn.~(\ref{eqn:finite-size-scaling}), we can obtain $\Pi_{\rm{gas}}^{\infty}$ and $\Pi_{\rm{liq}}^{\infty}$ as
\begin{equation}
    \Pi_{\rm{gas}}^{pa}=0.09163\pm0.0014;{\quad}\Pi_{\rm{liq}}^{pb}=0.1489\pm0.0088
    \label{eqn:percolation-pigas}
\end{equation}
The percolation thresholds are comparable to the bond percolation and site percolation thresholds of a pruned Voronoi network~\cite{jerauld1984percolation}. The percolation threshold of liquid-like droplets is larger than that of gas-like bubbles. Again, this result indicates that the percolation of gas-like bubbles can occur at lower concentration of gas-like molecules since the gas-like molecules have large local Voronoi volumes which enable them to be easily connected with each other throughout the entire system. 

\subsection{Location of the supercritical gas-liquid coexistence region}
\begin{figure}
	\includegraphics{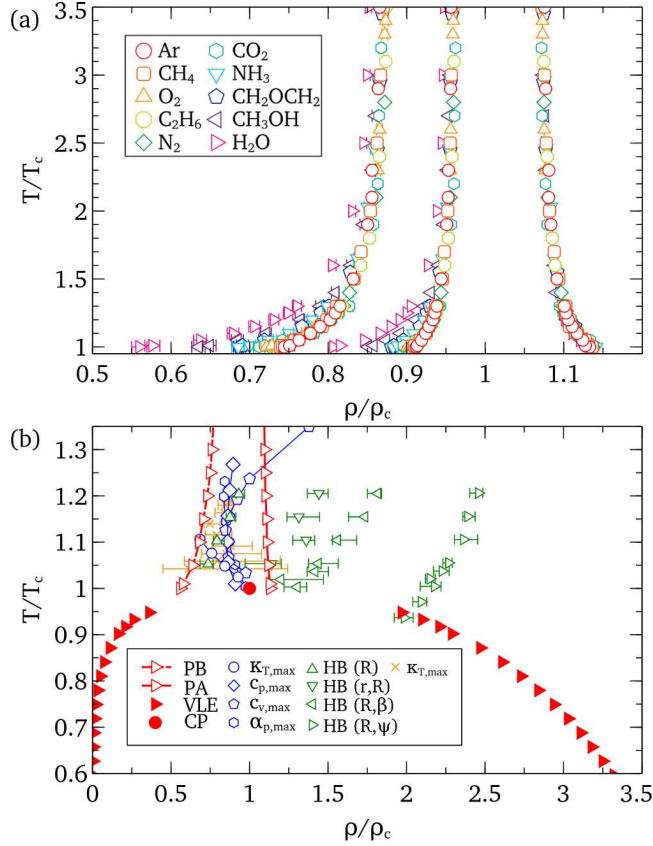}
    \caption{(a) Supercritical gas-liquid coexistence regions of various substances on the $\rho_{r}$-$T_{r}$ diagram. From left to right, the three branches correspond to $\rho_{pb}$, $\rho_{sgl}$, and $\rho_{pa}$, respectively. The structural transition region of non-polar substances almost coincide with each other. The transition region of strongly polar substances deviate from that of non-polar species near $T_{c}$. The error bars indicate the confidence interval of the transition densities estimated from the Eqn. (\ref{eqn:percolation-pigas}). (b) Comparison of the supercritical gas-liquid coexistence region of TIP4P/2005 water model, the ridges of maxima~\cite{schienbein2018investigation,strong2018percolation}, and the hydrogen percolation lines obtained from different criteria~\cite{strong2018percolation}. Here, $\mbox{R}, \mbox{r}, \beta, \psi$ correspond to the $\mbox{OO}$ distance, intermolecular $\mbox{OH}$ distance, $\angle{OOH}$ angle and the angle between the intermolecular $\mbox{OH}$ vector and the H-bond acceptor molecule normal, respectively.} The vapor/liquid equilibria data are from the work of Vega, Abascal, and Nezbeda~\cite{vega2006vapor}. The ridges of maxima and the hydrogen bonding percolation lines were read using a graph digitizer.
    \label{fig:rhoT}
\end{figure}
\begin{figure}
	\includegraphics{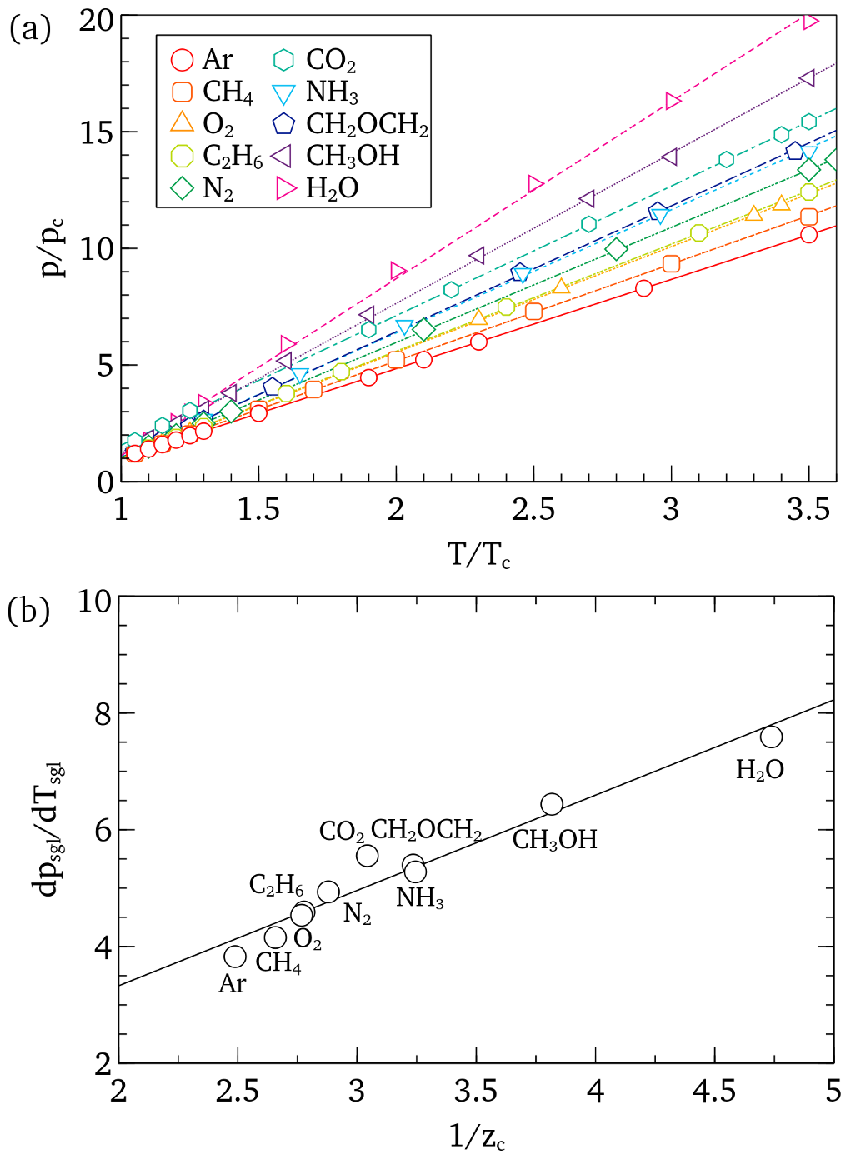}
    \caption{(a) Supercritical gas-liquid boundaries ($\Pi_{\rm{gas}}=0.5$) on the reduced pressure-temperature diagram. (b) Correlation between the slope of the supercritical gas-liquid boundary and the critical compressibility factor($z_{c}$). The slope is proportional to the inverse of $z_{c}$.}
    \label{fig:pTdiagram}
\end{figure}
Since we obtain the percolation thresholds of gas-like bubbles and liquid-like droplets for an infinite system, we can calculate the percolation transition densities $\rho_{pa}$ and $\rho_{pb}$ by solving Eqn.~(\ref{eqn:pigas-sigmoid}) for $\rho$. In addition, we can calculate the supercritical gas-liquid boundary as a set of densities where the gas-like fraction becomes 0.5 from Eqn. (\ref{eqn:pigas-definition}) ($\rho_{sgl}=\log(a)/b$). Hence, the Widom delta, the supercritical gas-liquid coexistence region which is enclosed with two percolation transition loci and bisected by the supercritical gas-liquid boundary, is completely defined from the gas-like fraction data.

Fig. \ref{fig:rhoT}{a} shows the supercritical gas-liquid coexistence region on the $\rho_{r}$-$T_{r}$ diagram. As expected from the $\Pi_{\rm{gas}}$ curves, the crossover densities ($\rho_{sgl}$, $\rho_{pa}$ and $\rho_{pb}$) of different substances mostly agree with each other except for strongly polar molecules including water, ethylene oxide, ammonia, and methanol near the critical temperature. In the case of methanol and water, $\rho_{pb}$ is lower than that of non-polar substances, whereas $\rho_{pa}$ lines agree with the others. This discrepancy would reflect the asymmetry of the influence of the attractive force on the structure of supercritical fluid. In contrast to dilute gas, the structural characteristics of dense supercritical fluids are mainly determined by the repulsive forces~\cite{weeks1971role,toxvaerd2015role}. Since the steepness of the repulsive part of the LJ potential is not relevant to the magnitude of the attractive interaction, $\rho_{pa}$ does not show a significant dependence on the type of the substances. In other words, $\rho_{pa}$ determines the boundary where the structure of supercritical fluid can be interpreted without consideration of the attractive force. On the other hand, $\rho_{pb}$ shows a substance-dependence because the influence of the attractive force becomes significant in the low-density region.

Of special interest is the case of water with hydrogen bonding network. Fig. \ref{fig:rhoT}b compares the supercritical gas-liquid coexistence region of water with the ridges of the response function maxima~\cite{schienbein2018investigation,strong2018percolation}, and the hydrogen bonding percolation lines~\cite{strong2018percolation}. The ridge of the compressibility maxima obtained from the E3B3 model \cite{strong2018percolation} is included in the supercritical gas-liquid coexistence region. Moreover, the ridges of isothermal compressibility ($\kappa_{T,max}$), isobaric heat capacity ($c_{p,max}$), and thermal expansion coefficient ($\alpha_{p,max}$) maxima on isotherms determined from the IAPWS95 equation of state~\cite{wagner2002iapws} are within supercritical gas-liquid coexistence region in spite of the incompleteness of the TIP4P/2005 forcefield~\cite{shvab2015thermophysical}. The line of the isochoric heat capacity maxima ($c_{v,max}$), which is known to show an oscillatory behavior due to the limit of the equation of state and terminate at a certain thermodynamic state, deviates from the Widom delta at higher temperatures. Note that hydrogen-bond percolation lines show different behaviors depending on the Hydrogen-bond criteria\cite{strong2018percolation}. One of the percolation lines, which only considers the distance $R$ between oxygen atoms of water molecules, is within the transition region, but the others which consider either the intermolecular $\mbox{OH}$ distance or the orientation of water molecules are away from the supercritical gas-liquid coexistence region. This result indicates that the response function maxima and the supercritical gas-liquid coexistence region are mostly relevant to spatial heterogeneity; note that the orientational anomalies discovered in the water systems were observed far below the critical density \cite{hestand2019mid,skarmoutsos2017local,ma2011density}.

Next, we explore the pressure-temperature relation. As shown in our earlier work on the Lennard-Jones particle \cite{yoon2018probabilistic,ha2018widom}, they form a deltoid region regardless of the substances (see the numerical data in the Supplementary material). This linear relation is surprising since the classification algorithm does not consider any thermodynamic variables to classify a molecule; it only exploits the geometrical characteristics of the Voronoi cells and their relationships with neighbors. Fig. \ref{fig:pTdiagram}a shows the supercritical gas-liquid boundaries of various substances. The transition pressures at which $\Pi_{gas}$ becomes 0.5 linearly depend on the temperatures, but their slopes are different from each other. Non-polar molecules have lower slopes than polar or hydrogen-bonded ones. This substance dependence of the transition pressures can be understood based on the following relation. The pressure-temperature relation is formulated as:
\begin{equation}
    p_{r}v_{r}=\frac{zT_{r}}{z_{c}}; \frac{p_{r}}{T_{r}}=\frac{\rho_{r}z}{z_{c}}
    \label{eqn:p-T-relation}
\end{equation}
where $z_{c}$ is the critical compressibility factor. By analyzing the limiting behavior of the pressure-temperature relation at high temperature ($T\rightarrow\infty$), we can derive a simple expression from Eqn. (\ref{eqn:p-T-relation}) to understand the pressure-temperature relation of supercritical gas-liquid boundary. As shown in Fig. \ref{fig:rhoT}, the reduced crossover densities ($\rho_{r,sgl}$) converge to 0.958 at high temperature regardless of the substances. Provided that the linear relation between the pressure and the temperature holds at the temperature limit, we can obtain the following equation.
\begin{equation}
\frac{p_{r,sgl}}{T_{r,sgl}}=\frac{0.958}{z_{c}}{\quad}\mbox{as}{\quad}T\rightarrow\infty
\end{equation} 
Fig. \ref{fig:pTdiagram}b shows that this simple relation can be regarded as a good approximation. The slope of the supercritical gas-liquid boundary is well expressed as a linear function of the compressibility factor ($dp_{r,sgl}/dT_{r,sgl}=1.633z_{c}^{-1}+0.061$). Therefore, the supercritical gas-liquid boundary can be expressed as:
\begin{equation}
p_{r,sgl}=(1.633z_{c}^{-1}+0.061)(T_{r,sgl}-1)+1
\end{equation}
This linear relation also holds for the percolation transition pressures. Hence, this result shows that the fractions of gas-like molecules (gas-likeness) of two substances are equal to each other at the same reduced pressure and temperature if their compressibility factors are equal. Given that the compressibility factor is a function of the acentric factor ($\omega$)~\cite{kulinskii2013critical}, this result also implies that the supercritical gas-liquid transition follows the three-parameter corresponding state principle, which states that all fluids with the same value of $\omega$ (or $z_{c}$) will have the same fraction of gas-like molecules at the same $T_{r}$ and $p_{r}$. That is, supercritical gas-liquid coexistence regions of two different substances on the $p_{r}-T_{r}$ diagram agree with each other if their critical compressibility factors (or acentric factors) are the same.

\section{Conclusions}
In conclusion, the classification algorithm based on the radical Voronoi tessellation in conjunction with the percolation theory reveals the generality of the Widom delta among simple molecular fluids. The fraction of gas-like molecules does not show significant substance-dependence except for strongly polar or hydrogen bonding molecules near the critical temperature. At $T{\gg}T_{c}$, the percolation transition lines and the supercritical gas-liquid boundary of all substances including the hydrogen bonded ones collapse to single lines. When the percolation probabilities are expressed as functions of $\Pi_{\rm{gas}}$, they collapse into a single line. A finite-size scaling can successfully locate the percolation thresholds of gas-like clusters (bubbles) and liquid-like clusters (droplets). By comparing the supercritical gas-liquid coexistence region of water and the lines of response function maxima, we demonstrate that the Widom delta include most of these lines. This result indicates that the notion of the supercritical gas-liquid coexistence region is suitable to describe the anomalous behavior of supercritical fluids since it does not violate the continuity picture proposed by van der Waals as well as connects the structural characteristics with the thermodynamic anomalies. The linear relationship between the reduced pressure and the reduced temperature observed in the Widom Delta of the monatomic Lennard-Jones system still holds for those of all substances studied in this work. The supercritical gas-liquid boundary and the percolation transition lines follow the three-parameter corresponding state theorem. Noting that the near-critical heat capacity maxima can be framed into the similarity law~\cite{banuti2017similarity}, the generality of the supercritical gas-liquid coexistence region implies that a general two-state model~\cite{russo2018water} would be possibly constructed to link the relation between the response functions and the structural characteristics. Hence, the algorithm designed in this work would be helpful to build a structure-property relation to understand the thermophysical behavior of supercritical fluids.

\section*{Supplementary material}
In supplementary material, we provide the numerical data that can help understand and reproduce the results in the main article. It includes the algorithm validation result; pressure data of near-critical fluids used for the estimation of the critical point; the fractions of gas-like molecules; the probabilities of finding an infinite droplet (bubble); the percolation transition densities (pressures); the number of independent gas-like bubbles and liquid-like droplets in $N=1,000$ systems; and the finite-size scaling data.

\section*{Acknowledgements}
M.Y.H. and W.B.L. acknowledge the support of National Research Foundation of Korea Grants funded by Korean Government (NRF-2015R1A5A1036133, NRF-2017H1A2A1044355).

\bibliography{bibfile}
\end{document}